\documentclass[twocolumn,twoside,slac_two]{revtex4}
\usepackage{graphicx}
\usepackage{fancyhdr}
\begin{document}

\title{New results and Possibilities in K Physics}

%

\author{T. Inagaki}
\affiliation{Institute of Particle and Nuclear Studies, High Energy Accelelator Research Organization (KEK) 1-1 Oho, Tsukuba Ibaraki 305-0801 Japan}

\begin{abstract}
Several topics in the K physics, $\left| V_{ud}\right|$ measurements, a check of $\mu$-e universality, and new CP-violation searches, are introduced. 
On the experimental plans of $K \to \pi\nu \bar{\nu}$ measurement using $K^+$ and $K_L^0$, their basic method and present status are discussed. 
We also make a very brief introduction of the new facility, J-Parc.
\end{abstract}

\maketitle

\thispagestyle{fancy}


\section{Introduction}
The flavor physics and CP violation, 
which is the main theme of B physics,
 has been also investigated in K physics for a long time. The first discoveries of particle-antiparticle ($K^0$-$\bar{K^0}$) mixing and CP violation are examples of its achievements. Since K contains s-quark, 
the K-physics is complementary with the B-physics which studies the transitions from b-quark. 

In an experimental viewpoint, most of K experiments have been performed as fix-target experiments in contrast with the fact of mostly collider experiments in the case of B. They can collect a large amount of decays to make highly precise measurements, or to attack very rare modes.
\par
The biannual international meeting of K-physics, KAON07, will be held at Frascatti next week.  Here, several recent topics are picked up from the KAON07 program: $\left| V_{us}\right|$ measurements, check of $\mu$-e universality through $K_{l2}$ decays, new CP-violation searches through the charge asymmetry in $K_{\pi 3}$ decay and muon transverse polarization in $K_{\mu 3}$ decay and $K \to \pi \nu \bar{\nu}$ experiments using $K^+$ and $K_L^0$.
For other topics and more detail information about recent K-physics, please visit the web site of KAON07. 
\cite{kaon07}

\section{$\left|V_{us}\right|$ measurements}
$\left| V_{us}\right|$, $\lambda$, is  a critical ingredient in a determination of the other CKM parameters as seen in the Wolfenstein parameterization below.

\small{
\begin{eqnarray}
\lefteqn{V_{CKM}  =  \left(
\begin{array}{cccc}
V_{ud}& V_{us}& V_{ub} \\
V_{cd}& V_{cs}& V_{cb} \\
V_{td}& V_{ts}& V_{tb}
\end{array}
\right)} \\
&& =  \left(
\begin{array}{cccc}
1-\lambda^2/2& \lambda& A\lambda^3(\rho-i\eta) \\
-\lambda& 1-\lambda^2/2& A\lambda^2 \\
A\lambda^3(1-\rho-i\eta)& -A\lambda^2& 1 \\
\end{array}
\right) + O(\lambda^4) \nonumber
\label{eq-ckm}
\end{eqnarray}
}

The precise determination of $\left|V_{us}\right|$ is also critical to check the untarity of the CKM matrix using its first row. In the review of PDG-2004 there was a hint of deviation from unitarity by 2.2$\sigma$ as shown in  Eq.~\ref{eq-unitality1} 

\begin{equation}
1-(\left|V_{ud}\right|^2+\left|V_{us}\right|^2+\left|V_{ub}\right|^2)=(3.1\pm1.5)\times10^{-3} 
\label{eq-unitality1}
\end{equation}

\begin{figure}[h]
\centering
\includegraphics[width=80mm]{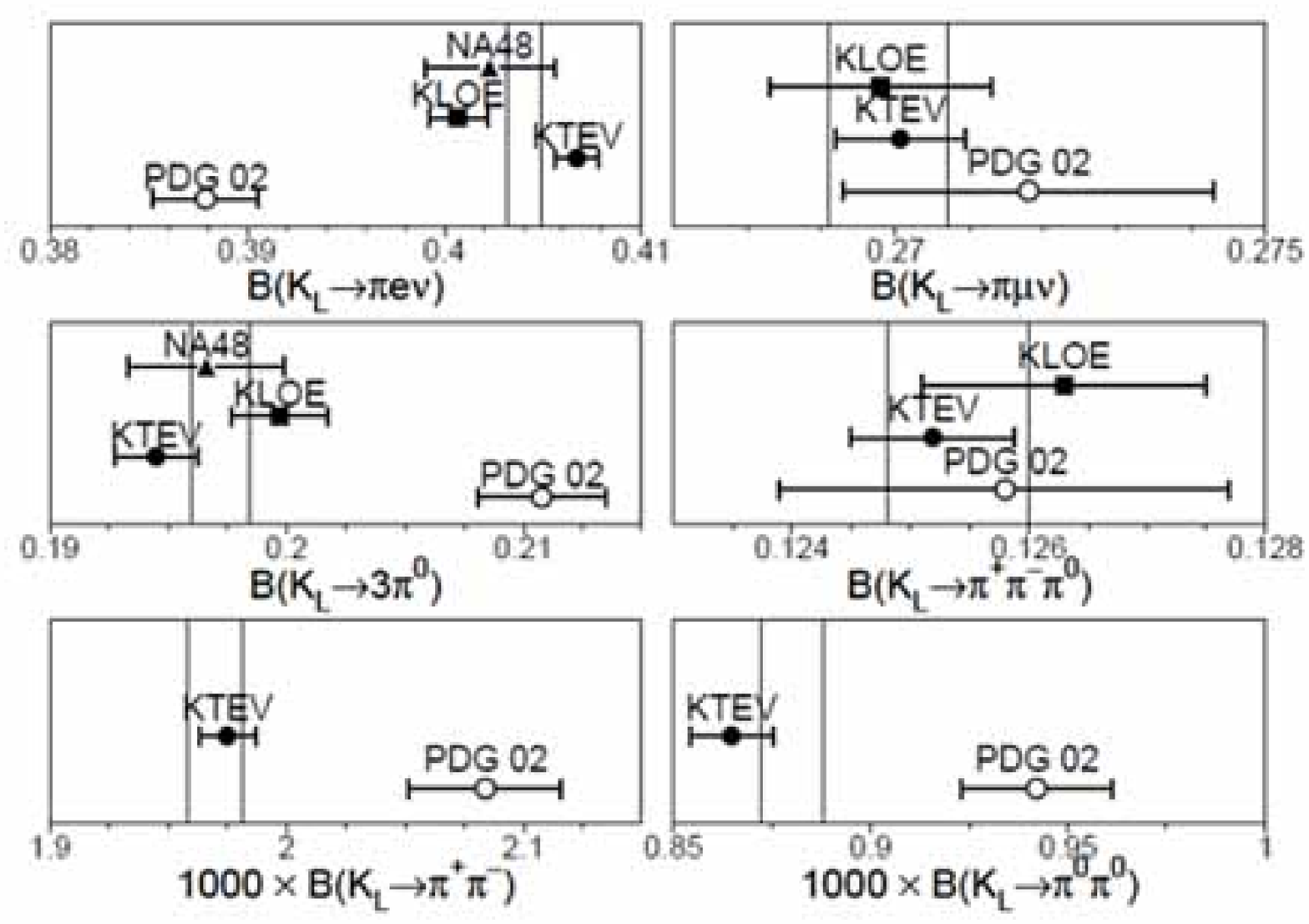}
\caption{Branching ratios of $K_L^0$ decays. The vertical lines indicate the $\pm 1\sigma$ bounds from a fit to all KTeV, KLOE and NA48 measurements. Figures were copied from Ref\cite{blucher}} 
\label{figure1}
\end{figure}

The blanching ratios of all $K_{l3}$ decays together with their form factors for $K_L^0$ and $K^+$ were re-measured by KTeV, KLOE, NA48 and ISTRA.  The life time of $K_L^0$ and branching ratios of $K_{e3}$ of $K_S^0$ and $K_{\mu2}$ were also re-measured by KLOE. Meanwhile, most of branching ratios of K decays have been revised and significantly changed from the previous values. Figure~\ref{figure1} shows $K_L^0$ branching ratios as an example.

One of reasons of the changes was suggested by KTeV as it was caused with a previous inadequate treatment of inner bremsstrahlung. Now, $\left|V_{us}\right|$ is determined with an accuracy of less than 1 \%.
\par
The $\left|V_{ud}\right|$, which is a unique counterpart for the unitarity check because of the smallness of $\left|V_{ub}\right|$, has been determined with an improved accuracy by a factor of 2 through a global study of nine super-allowed beta-decays and a refined radiative correction. Then, the unitarity becomes as shown in  Eq.~\ref{eq-unitality2}. 
\cite{palutan}
\begin{equation}
1-(\left|V_{ud}\right|^2+\left|V_{us}\right|^2+\left|V_{ub}\right|^2)=(0.2\pm1.4)\times10^{-3} 
\label{eq-unitality2}
\end{equation}

Although the unitarity is recovered with a slightly better accuracy, the precision of $\left|V_{us}\right|$ determination is worse than that of $\left|V_{ud}\right|$.  It is certain that the more improvement of $\left|V_{us}\right|$ determination is required in order to check the unitarity the more critically. 
\par
One of important aspects, which should be stressed, is an influence or a stimulation to other measurements and related physics by the recent efforts to improve $\left|V_{us}\right|$ and  $\left|V_{ud}\right|$ determinations. The ratio of $\tau$-decays into hadrons with those into leptons turns to be a channel to determine the mass of s-quark using the improved $\left|V_{us}\right|$. Measurements of neutron life time and $g_A/g_V$, which must be the best for $\left|V_{ud}\right|$ determination later and which closely relate to nucleo-synthesis scenario, are also being stimulated. The recent works have been achieved by various theoretical studies as well as the experimental efforts, but there still remains further theoretical studies, such as the improvements of estimation of $f_K$ and $f_K/f_\pi$. They are the parameters bringing the largest ambiguity to the determinations $\left|V_{us}\right|$ from $K_{l3}$ and $K_{l2}$, respectively.  The studies are valuable not only for the next $\left|V_{us}\right|$ determination but also for a development of the method to be applied to other physics.

\section{Check of $\mu$-e universality}
Among various checks of the $\mu$-e universality, the ratio of Br($K^+ \to e^+ \nu$)/Br($K^+ \to \mu^+ \nu$) (denote it $K_{e2}/K_{\mu 2}$) has an advantage. Most of theoretical ambiguities in each
individual decay, for example the decay constant $f_K$, are cancelled by taking ratio. The standard model predicts the ratio to be $(2.472\pm 0.001)\times 10^{-5} $. The ambiguity is only 0.04 \%. On the other hand, the experimental upper bound is $(2.45\pm 0.11)\times 10^{-5} $, the error of which is about 4 \%. There is enough room to look for new physics. Recently, we have heard interesting progresses in both theory and experiment.
\par
A recent calculation based on SUSY with lepton flavor violation mechanism by Masiero {\it et al.} \cite{masiero} predicts that the effect can be as high as 2 \%, which is almost near to the experimental limit. In the model, right-handed neutrino, which is a major object in the present particle physics, plays a crucial role.
\par
On the experimental side 
\cite{wanke}
two runs of NA48-2 and KLOE recently improved the accuracy by a factor of 2 individually and the combined error reached 1.3 \%. Moreover, the NA48 collaboration announced that they have a chance of new measurement using the beam from this July during the test run for their next plan of $K^+ \to \pi^+ \nu \bar{\nu}$ measurement (P326, NA48-3). They claimed that their major systematic error in NA48-2 was background contamination to $K_{e2}$ by $K_{\mu 2}$ and that it can be controlled in the new measurement.  The goal of the new measurement is $\pm 0.3 \%$. 
\par
On the other hand, the error of $K_{e3}/K_{\mu 3}$ was also reduced from $1 \%$ to $0.5 \%$ using many data for  $\left|V_{us}\right|$ measurements. However, the theoretical ambiguity, mostly from form factor difference of $K_{e3}$ and $K_{\mu 3}$, is dominant.  Another check of $\mu$-e universality using pion decay, PIENU at TRIUMF, plans to improve the ratio of Br($\pi^+ \to e^+ \nu$)/Br($\pi^+ \to \mu^+ \nu$) by a factor of 5.

\section{New CP-violation searches} 

\subsection{Charge asymmetry in $K_{\pi 3}$ decays}
The charge asymmetry in $K_{\pi 3}$ decays, A$_g$, is suppressed to be around $10^{-5}$ in the standard CKM model and the previous limit of experiment was an order of $10^{-3}$. NA48-2 measured the charge asymmetry for $K^{\pm}\to \pi^{\pm} \pi^0 \pi^0$ (A$_g^0$) and that for $K^{\pm}\to \pi^{\pm} \pi^+ \pi^-$ (A$_g^c$) in order to look for a new CP-violation source.
\cite{ichep06} Using hundreds million samples of both decays, they obtained the results as A$_g^0 = (2.1 \pm 1.9)\times10^{-4} $ and A$_g^c = (-1.3 \pm 1.7)\times10^{-4} $. The values are consistent with zero, and they could set new upper bounds, which were one-order-of-magnitude below the previous ones.  Although some room of another one-order remains down to the standard model prediction, it may be said that the accuracy reached to the level that no one wants to re-measure in near future. It was one of the best measurements in K, in which systematic errors were well controlled.
 
\subsection{Transverse muon polarization in $K_{\mu 3}$ decay}
\begin{figure}[h]
\centering
\includegraphics[width=60mm]{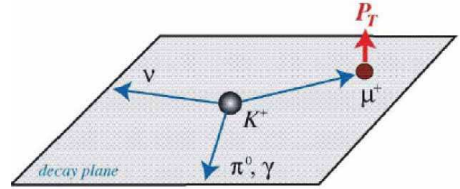}
\caption{Transverse muon polarization in $K_{\mu 3}$ decay. $P_T$ is defined as $\frac{\sigma_{\mu} \left(P_{\pi^0}\times P_{\mu^+}\right)}{\left|P_{\pi^0}\times P_{\mu^+}\right|}$} 
\label{figure2}
\end{figure}

Another search of a new CP-violation source is the transverse muon polarization ($P_T$) in $K^+\to \pi^0 \mu^+ \nu$ decay as shown in Figure~\ref{figure2}.

Non-zero $P_T$ is a signature of T-violation, which corresponds to CP-violation under CPT theorem. The standard CKM model predicts $P_T$ to be less than $1\times10^{-7}$, but a spurious t-odd effect due to final state interactions is expected to appear at $1\times 10^{-5}$, which determines the limit of search.  Recently, E246 collaboration at KEK reported their final results.\cite{e246}
  They well controlled most of sources of systematic error by using stopped $K^+$, a 12-gap toroidal spectrometer and symmetrically arranged $\pi^0$ detectors.
%
  
The result was $P_T = -0.0017 \pm 0.0023(stat) \pm 0.0011(syst)$. It was consistent with zero and they set an upper bound of $P_T < 0.0050$. The group is proposing a new experiment (TREK) at J-Parc to go down to $1\times 10^{-4}$. They clain that the largest systematic error, which came from an ambiguity of magnetic field at the muon stoppers ($\mu$SR position), will be reduced by replacing the stopper from aluminum plates to active ones. The collaboration will re-use most of the E246 detectors with small modifications. The proposal passed the first step examination.

\section{Status of the experiments for charged and neutral modes of $K \to \pi \nu \bar{\nu}$ decay}

The $K \to \pi \nu \bar{\nu}$ decay has been considered as a golden-plated channel in K-decays and now there are three plans; P326 (NA48-3) at CERN-SPS for the charged mode $K^+\to \pi^+ \nu \bar{\nu}$, and KLOD at Serpkov-U70 and E14 at JPARC-PS for the neutral mode $K_L^0 \to \pi^0 \nu \bar{\nu}$.  P326 and E14 are taking a similar strategy of step-by-step approach to reach an observation of 100-events (for the standard model predictions). 

\par
\begin{figure}[h]
\centering
\includegraphics[width=60mm]{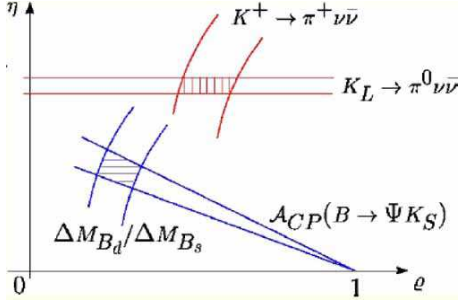}
\caption{One possible scenario: Difference of CKM parameters determined by K and B} 
\label{figure4}
\end{figure}

In the standard model the $K^+\to \pi^+ \nu \bar{\nu}$ decay contains information of  $\left|V_{td}\right|$ with adding a small charm-quark contribution, and the $K_L^0 \to \pi^0 \nu \bar{\nu}$ decay does Im$V_{td}$. The branching ratios, which are predicted by the standard model using the CKM parameters determined by B-decays, are $(7.8 \pm 1.2) \times 10^{-11}$ for $K^+$ and $(3.0 \pm 0.6) \times 10^{-11}$ for $K_L^0$. It is quite interesting that the errors already reached $\pm 15 \%$ and $\pm 20 \%$, respectively. So, we may say that the main goal of the experiments is not additional measurements of  $\left|V_{td}\right|$ or Im$V_{td}$. It is a search of a deviation due to new physics as shown in Figure~\ref{figure4}.

The $K \to \pi \nu \bar{\nu}$ decay has remarkable merits in the search. The decay proceeds through purely electro-weak FCNC (Flavor Changing Neutral Current) and the theoretical ambiguity is very small. It is a few $\%$. For example, a measurement of the $K_L^0 \to \pi^0 \nu \bar{\nu}$ decay at a sensitivity of $ 3\times 10^{-13}$ corresponds to the observation of 100 SM-events and corresponds to an observation of a $5\sigma$ effect for the new physics, which enhances the branching ratio by a factor of 1.75. The 1.75 enhancement can be brought by physics of 100 TeV mass scale in a simple tree diagram model. In the case of $B \to X_s \mu^+ \mu^-$ and $B_s \to \mu^+ \mu^-$, which are considered as golden plated channels in new-physics searches in B-physics at LHC, the 1.75 enhancement corresponding observations of 100 SM-events and $5\sigma$ effect can be brought by 10-TeV scale in the similar model. Such an example varies model-by-model, but it indicates that the measurement might be still important even in the era of LHC.     

\subsection{$K^+\to \pi^+ \nu \bar{\nu}$ decay}

\begin{figure}[h]
\centering
\includegraphics[width=70mm]{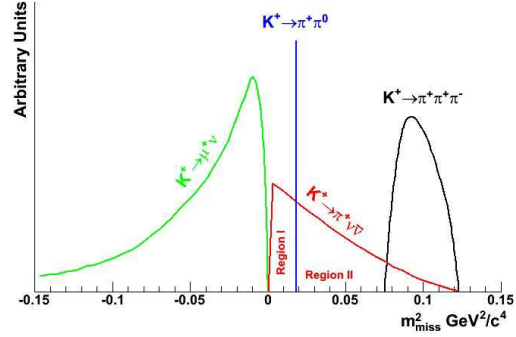}
\caption{Expected $m_{miss}$ distribution} 
\label{figure5}
\end{figure}

 The basic method taken by P326 
\cite{p326}
is a missing mass method by measuring the momenta of incident and secondary particles, which are $K^+$ and $\pi^+$ for the $K^+\to \pi^+ \nu \bar{\nu}$ decay. The missing mass ($m_{miss}$) can be calculated from Eq.~\ref{pinunubar-mm}. 

\begin{equation}
m_{miss}^2 \approx m_K^2 \left[1-\frac{\left|P_{\pi}\right|}{\left|P_K\right|}\right] + m_{\pi}^2 \left[1-\frac{\left|P_K\right|}{\left|P_{\pi}\right|}\right]
-\left|P_K\right| \left|P_K\right| \theta^2,  
\label{pinunubar-mm}
\end{equation}

where $\theta$ is an angle of the $\pi^+$ direction with respect to the $K^+$ direction. The $m_{miss}$ distribution is expected as shown in Figure~\ref{figure5}.

The $K^+\to \pi^+ \nu \bar{\nu}$ decay can be found in the two valleys under three bumps or peaks from decays of $K_{\mu2}$, $K_{\pi2}$ and $K_{\pi3}$. The amount of events in each bump is expected to be $10^{10}$ times larger than that of $K^+\to \pi^+ \nu \bar{\nu}$. Therefore, it is crucially important to measure both momenta as well as tracks to determine the vertex accurately as much as possible. In addition to the missing mass, they will install PID (RICH detector) to reduce $K_{\mu2}$ and photon veto system to reduce $K_{\pi2}$. Still, the keenest element is the tracker for incident $K^+$, which should be operated under a high beam-rate of 800 MHz.
The estimated values of signal and background are listed in ~Table ~\ref{table1}.

\begin{table}[h]
\begin{center}
\caption{Signal and background expected in P326, The Region 1 and 2 are the regions where $m_{miss}$ stays in the valley below and above $K_{\pi2}$, respectively.}
\begin{tabular}{|l|c|c|c|}
\hline \textbf{Events/year} & \textbf{Total} & \textbf{Regin 1} &
\textbf{Regin 2} \\ 
\hline\hline Signal & 65 & 16 & 49 \\
\hline\hline $K^+\to \pi^+ \pi^0$ & 2.7 & 1.7 & 1.0 \\
\hline $K^+\to \mu^+ \nu$ & 1.2 & 1.1 & <0.1 \\
\hline $K^+\to e^+ \pi^+ \pi^- \nu$ & 2 & negligible & 2 \\
\hline Other 3-track decays & 1 & negligible & 1 \\
\hline $K^+\to \pi^+ \pi^0 \gamma$ & 1.3 & negligible & 1.3 \\
\hline $K^+\to \mu^+ \nu \gamma $ & 0.5 & 0.2 & 0.2 \\
\hline $K^+\to e^+ (\mu^+) \pi^0 \nu$, others & negligible & - & - \\
\hline\hline Total background & 9 & 3.0 & 6 \\
\hline
\end{tabular}
\label{table1}
\end{center}
\end{table}
 
They claimed to be able to collect 65 events /year with an S/N of around 8. 
\par

P326 is a successor of NA48, as sometimes called as NA48-3. They will be able to re-use major parts of NA48 setup as well as their rich experience. The collaboration is waiting for a financial support for the detector modification and they will start from a beam test run in this July as mentioned in the previous section.

\subsection{$K_L^0\to \pi^0 \nu \bar{\nu}$ decay}

In the neutral mode, there are no $K_{\mu2}$ and $K_{\pi2}$ branching ratio is smaller than that in the charged mode by three orders of magnitude. Although they are great advantages, The $K_L^0\to \pi^0 \nu \bar{\nu}$ decay has several disadvantages, such as relatively poor resolution for a clean separation of the regions of signal and background and no tracking devices of photons. Then, all planned experiments use a very thin beam (pencil beam) to identify the $K_L^0\to \pi^0 \nu \bar{\nu}$ decay with a missing transverse-momentum ($P_T$) method. The energies and hit positions of two photons from $\pi^0$ are measured with calorimeter, the decay vertex ($Z_{vtx}$) is calculated with an assumption that the two-photons are created from $\pi^0$ on the beam axis, and $P_T$ is calculated by using the decay vertex. Contaminations by other decays are reduced by veto cut. 
\par
\begin{figure}[h]
\centering
\includegraphics[width=80mm]{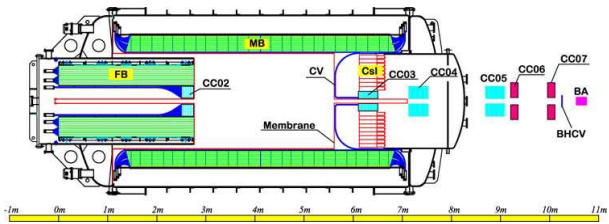}
\caption{Experimental setup of E391a. It will be used for the E14 experiment at J-Parc with modifications in some parts.} 
\label{figure6}
\end{figure}

For a further pursuit beyond the previous KTeV trial, there were many technical problems such as construction of the pencil beam, a very high vacuum of $10^{-5}$ Pa to reduce $\pi^0$ productions in the decay region by beam-air interactions, very tight veto with a threshold down to 1 MeV to reduce contaminations by $K_{\pi2}$, {\it etc}, and a fundamental question whether the method can work or not. E391a at KEK-PS was planned and carried out to get answers to these problems and question by looking at a final plot in a  high sensitivity.
Figure~\ref{figure6} shows the experimental setup of E391a. The detectors, which were cylindorically arranged, were installed in a vacuum vessel.

\begin{figure}[h]
\centering
\includegraphics[width=70mm]{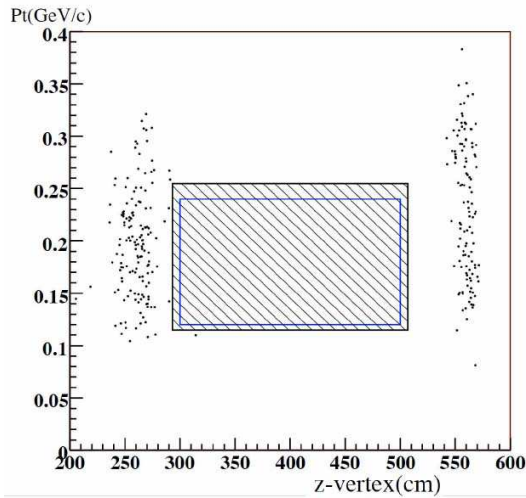}
\caption{$Z_{vtx}$ vs $P_T$ plot for the Run-2 1/3 sample} 
\label{figure7}
\end{figure}

Figures~\ref{figure7} shows a preliminary plot of $Z_{vtx}$ vs $P_T$ using a partial data sample.\cite{sample} 

Although the signal box is still masked because of on-going studies for a so-call blind analysis, there is no event in a surrounding region except for the up- and down-stream $Z_{vtx}$, which correspond to the positions of the collar counter CC02 and the veto counter for charged particles. Those counters have a small aperture for beam, in other word, they provided materials placed near the beam axis. The events clusters are consistent with $\pi^0$ produced by interactions with beam halo. 
%

\par
\begin{figure}[h]
\centering
\includegraphics[width=60mm]{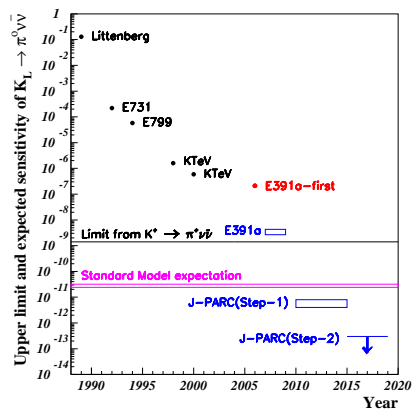}
\caption{Achievements expected in a series of experiments in KEK and J-Parc} 
\label{figure9}
\end{figure}

The collaboration also takes a step-by-step approach at J-Parc. 
\cite{e14}
The step-1, E14, is a plan to reach a sensitivity of $O(10^{-12})$ and the step-2 $O(10^{-13})$. E14 also passed the first examination. ({\it E14 was fully-approved in July and it goes into a stage of preparation to catch up the first beam of J-Parc.}) The design concept of E14 is to recycle all possible devices of E391a, and to modify its parts, where modification is considered to be crucial for further background rejection and for the tolerance against the expected high-rate. The main modifications are a replacement of the CsI calorimeter into thick and finely-segmented one used in KTeV, a change of readout electronics into wave-form digitizer and a replacement of the beam plug counter to the one, which has been designed for KOPIO. The collaboration estimated that the contaminations by halo beam interactions can be reduce to a negligible level, using less halo and better K/n at the new beam line in J-Parc, optimizing the apertures and materials of collar counters, {\it etc}. These design studies were surely based on experimental data of E391a and experiences, and a similar communication is expected to happen between E14 and the step-2 design.  Figure~\ref{figure9} shows a biographical sketch of achievements expected in a series of experiments in KEK and J-Parc.  

\par
A measurement of $K_L^0\to \pi^0 \nu \bar{\nu}$, KLOD, is also planned at U-70 at Protovino.
\cite{klod}
Since one of E391a Russian members jointed to the design, the plan also follows the basic method of E391a.  They intend to construct a pencil beam line soon, however, detailed estimations of signal and background have not yet been reported to outside.
     
\section{Facilities for K-physics}

The K-physics programs using the US facilities, such as CPT, CKM and KAMI at Fermilab and E949 and KOPIO at BNL, were all cancelled in these few years.  Although the cancellations happened not only in K but also in all accelerator physics, the damage of K-physics was serious. 
\cite{usprogram}
Afterward, there remains a few facilities in the world: CERN-SPS, $\Phi$-factory, U-70 and KEK.
\par
\begin{figure}[h]
\centering
\includegraphics[width=80mm]{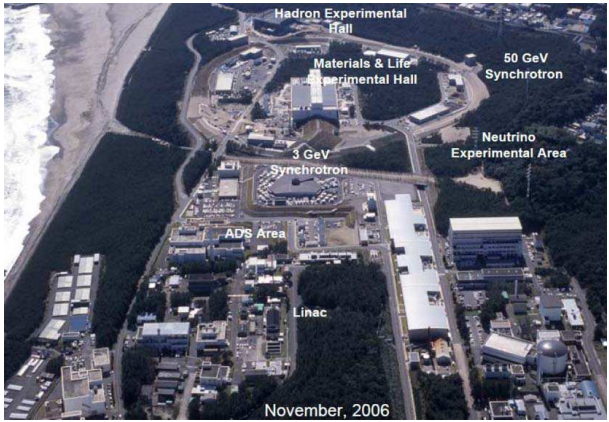}
\caption{Landscape of the J-Parc construction. Water in the left is the Pacific Ocean. } 
\label{figure11}
\end{figure}

Under such a tough situation we are really welcoming a new facility of J-Parc. The accelerator complex of J-Parc, composing of 3 GeV PS(proton synchrotron) and 50 GeV PS, was designed so as to feed highly intense beam of 1 MW to both fields: material science using spolation neutrons or muons and particle nuclear physics using neutrinos or K. The construction is about to reach the final stage. 
as shown in Figure~\ref{figure11}

The commissioning of 50 GeV PS is scheduled to be the end of 2008. Hopefully, J-Parc will open new frontier in both $\nu$ and K-physics, soon.

\bigskip 

\begin{thebibliography}{99}
\bibitem{usprogram} A possibility of recovery program was recently announced. It will be discussed at a special session of KAON07.
\end{thebibliography}

\end{document}